\documentclass{sig-alternate}
\usepackage{url}
\usepackage{caption}
\usepackage{verbatim}
\usepackage{multirow}
\usepackage{txfonts}
\usepackage{amsmath}
\usepackage{xfrac}
\usepackage{pslatex}

\begin{document}
%

\title{Fit or Unfit : Analysis and Prediction of `Closed Questions' on Stack Overflow}
%
%
%
%
%

\numberofauthors{2} 
\author{
\alignauthor
Denzil Correa, Ashish Sureka \\ 
\affaddr{Indraprastha Institute of Information Technology IIIT-Delhi }\\ 
\email{\{denzilc, ashish\} @iiitd.ac.in} \\ 
}

\maketitle
\begin{abstract}
Stack Overflow is widely regarded as the most popular Community driven Question Answering (CQA) website for programmers. Questions posted on Stack Overflow which are not related to programming topics, are marked as `closed'  by experienced users and community moderators. A question can be `closed' for five reasons -- \emph{duplicate, off-topic, subjective, not a real question} and \emph{too localized}. In this work, we present the first study of `closed' questions in Stack Overflow. We download 4 years of publicly available data which contains 3.4 Million questions. We first analyze and characterize the complete set of 0.1 Million `closed' questions. Next, we use a machine learning framework and build a predictive model to identify a `closed' question at the time of question creation. 

One of our key findings is that despite being marked as `closed', \emph{subjective} questions contain high information value and are very popular with the users. We observe an increasing trend in the percentage of closed questions over time and find that this increase is positively correlated to the number of newly registered users. In addition, we also see a decrease in community participation to mark a `closed' question which has led to an increase in moderation job time. We also find that questions closed with the \emph{Duplicate} and \emph{Off Topic} labels are relatively more prone to reputation gaming. For the `closed' question prediction task, we make use of multiple genres of feature sets based on - user profile , community process, textual style and question content. We use a state-of-art machine learning classifier based on an ensemble learning technique and achieve an overall accuracy of 73\%. To the best of our knowledge, this is the first experimental study to analyze and predict `closed' questions on Stack Overflow.

\end{abstract}




\section{Introduction}

\subsection{Research Motivation and Aim}

Community driven Question Answering (CQA) websites like Stack Overflow, Quora and Yahoo! Answers are popular contemporary genre of websites on the Internet. CQA websites follow a standard Q\&A format where a user asks a question on a problem she faces; while other users (who may have some prior expertise) respond with their answers on the question. Effectively, CQA websites follow a crowd sourced model in which the knowledge of experts is exploited to form a large scale knowledge base on variety of topics. Stack Exchange is a platform which provides libraries to deploy topic-based community powered Q\&A websites~\cite{Jeff-Atwood:2009fk}. The Stack Exchange platform is a growing network of CQA websites and currently supports 103 Q\&A forums on diverse topics like code review, parenting, bicycles and audio-video production\footnote{\scriptsize{\url{https://stackexchange.com}}}.

Stack Overflow is the first and most popular Stack Exchange website which caters to the benefits of professional programmers and programming enthusiasts.\footnote{\scriptsize{\url{http://stackoverflow.com}}} It is a free and open Q\&A website where users can ask programming related questions. Stack Overflow maintains a strong emphasis on question-answer based format of the site and strongly discourages discussion or \emph{chit-chat}. 
In particular, questions on the topics which contain specific programming problems, software algorithms, coding techniques and software development tools are recommended and considered fit for its Q\&A format. An intricate community based voting process is followed to reward users for good quality questions and answers. Relevant, technically challenging and good question-answers are rewarded by the community with \emph{votes}. Similarly, answers which address the problem encountered by the original question can be voted \emph{accepted}. This voting process allows post owners to earn a \emph{reputation} which is a reflection of their contribution worth to the Stack Overflow community. Conversely, the same voting process can lead to penalties on the post owner's \emph{reputation} due to low quality posts like wrong answers, spam and advertisements.~\emph{Badges} (the online equivalent of medals) are awarded to users as incentives to highlight special achievements based on community participation. This community based \emph{reputation reward} process helps to ensure a reasonable degree of high quality content on the website and weed out low quality content. 

\begin{figure}[ht!]
\captionsetup{font=footnotesize, labelfont=bf, textfont=bf}
\centering
\includegraphics[width=\linewidth]{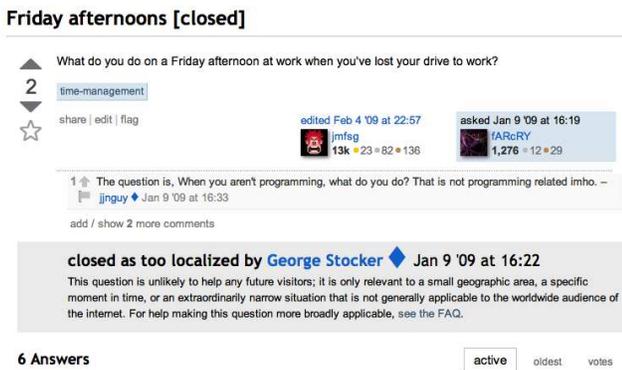}
\caption{shows a screenshot of question marked `closed' on Stack Overflow on account of being \emph{Too Localized}.}
\label{fig:clo-screenshot}
\end{figure}

Stack Overflow is a free, open website to all users and therefore, maintenance of content quality on such a large scale social collaborative platform is a challenge~\cite{agichtein2008finding}. Questions are an integral part of the Stack Overflow system and hence, quality control of questions play a significant role in its functioning and popularity. Stack Overflow guidelines clearly outline categories of questions which are deemed unfit for its Q\&A format. Opinion-based questions and questions which have a tendency to generate discussions rather than answers are categorically considered inappropriate. Some examples of such questions include (but not limited to) homework questions, product or service recommendations, non-programming related and polls. Questions on Stack Overflow which do not fall into one of the pre-defined set of guidelines are marked `closed' via a community-based voting system. A question can be marked as `closed' for five reasons -- \emph{duplicate, off-topic, subjective, not a real question} and \emph{too localized}. Section~\ref{sec:why-clo-q} contains a detailed discussion on the procedure to mark a question `closed' and its sub-categories. Figure~\ref{fig:clo-screenshot} shows an example of a `closed' question on Stack Overflow on account of being \emph{Too Localized}. 

A question is primarily marked `closed' either due to low quality or due to irrelevance to the Stack Overflow CQA platform. The decision to `close' a question lies completely on the shoulders of experienced users and community moderators via a systematic voting process. Due to exponential growth of Stack Overflow user base, there has been a steady increase in the workload on moderators. The process of marking a `closed' question also requires multiple context switches~\cite{:2013kx}. Despite the existence of vibrant experienced users and self-motivated community moderators, Stack Overflow faces a continuous ongoing challenge to maintain quality of questions on their website. Therefore, it is important to analyze and study the phenomena of `closed' questions in order to gain historical insights which can help make the future plan-of-action. 

The goal of Stack Overflow is to have a knowledge base of question-answers on programming related topics. A `closed' question is a direct feedback to the question asker that her question may be unfit or needs improvement in its current form. A system to predict a `closed' question at post creation time can serve as an early feedback mechanism on question quality to the question asker. Such a system would also help community moderators to identify and mark `closed' questions. Therefore, prediction of a `closed' question at post creation time has two distinct benefits -- (1) feedback to question asker and (2) community moderator assistance.


\subsection{Research Contributions}
We conduct the first study of `closed' questions on Stack Overflow.  We make the following research contributions --

\begin{itemize}
	\item We present a characterization of `closed' questions on Stack Overflow. We perform analysis on question content, answer patterns and temporal trend analysis of `closed` question. In addition, we also make observations on community participation trends towards `closed' questions as well as analyze information quality indicators on `closed' questions.

	
	\item We use an ensemble based machine learning framework to build a predictive model for `closed' question prediction on Stack Overflow and report 73\% accurate predictions. We analyze features and report top features to differentiate `closed' questions from non-`closed' questions.
	
\end{itemize}

To the best of our knowledge, this is the first focussed study on analysis and prediction of `closed' questions on Stack Overflow. The rest of the paper is organized as follows. Related work is covered in Section~\ref{sec:rel-work}. Section~\ref{sec:why-clo-q} gives an overview of `closed' questions, its sub-categories and procedures followed to `close' a question. Section~\ref{sec:clo-q-analysis} presents our characterization study of `closed' questions. Section~\ref{sec:clo-q-prediction} demonstrates a machine learning framework to predict `closed` questions on Stack Overflow and Section~\ref{sec:clo-q-conclusion} outlines Conclusion. 

\section {Related Work}
\label{sec:rel-work}

Stack Overflow is a popular Q\&A website used by programmers all over the world to seek answers to programming related questions~\cite{mamykina2011design}. Besides being a question-answer website, Stack Overflow has evolved into a knowledge base for programming related tasks~\cite{anderson2012discovering}. It has also been used for various other core and ancillary programming tasks like building crowd sourced API documentation, deficient documentation identification, identification of mobile development issues and improvement of bug tracking systems~\cite{campbell2013deficient,linares2013exploratory,416,parnincrowd}. Stack Overflow is a free and open website and has 1.29 Million registered users with 3.4 Million questions.\footnote{\scriptsize{as of August 2012}} Quality control of content on such large scale community driven collaborative systems is a research challenge~\cite{agichtein2008finding}. Questions and answers form an integral part of any CQA website and therefore, it is important to have quality checks in place for both questions and answers. 
 
Evaluation and prediction of answer quality has attracted wide spread attention in the IR research community. Jeon~\emph{et al.} propose maximum entropy and kernel density estimation approach in conjunction with non-textual features to  predict answer quality on Naver, a Korean CQA website~\cite{Jeon:2006:FPQ:1148170.1148212}. Agichtein \emph{et al.} model content quality in Yahoo! Answers with content and graph-based implicit user relationship features~\cite{agichtein2008finding}. Shah \emph{et al.} propose a classification model with features based on human assessed aspects and question-answer meta information to predict answer quality on Yahoo! Answers CQA~\cite{shah2010evaluating}. Sakai \emph{et al.} propose evaluation methods based on graded-relevance IR metrics to find the best answers on Yahoo! Chiebukuro (Japanese Yahoo! Answers)~\cite{sakai2011using}. All the above approaches focus on answer quality on large scale CQA websites. However, it is also important to focus on question quality because prior work shows that answer quality directly depends on question quality~\cite{agichtein2008finding}. Low quality questions have a direct impact on user experience, question retrieval, question recommendation and hence, it is important to maintain high question quality~\cite{mamykina2011design}. Li \emph{et al.} analyze factors affecting question quality and propose a Mutual Reinforcement-based Label Propagation approach to predict question quality in Yahoo! Answers~\cite{Li:2012:APQ:2187980.2188200}. To the best of our knowledge, this is the only work which addresses quality of questions on a large-scale CQA website. However, Stack Overflow is a programming related CQA and has appreciable differences in style, structure, organization and content from other generic CQA websites like Yahoo! Answers and Quora~\cite{mamykina2011design,wangwisdom}.

In context of all the previous work in this area, our study differs in two main aspects -- (1) we focus our attention on question quality rather than answer quality and (2) we perform our analysis on Stack Overflow which is a programming-based CQA. `Closed' questions on Stack Overflow are considered bad for its Q\&A format and hence, by definition are low quality given the context. 

\section{`Closed' Questions on Stack Overflow}
\label{sec:why-clo-q}

\begin{figure}[ht!]
\captionsetup{font=footnotesize, labelfont=bf, textfont=bf}
\centering
\includegraphics[width=\linewidth]{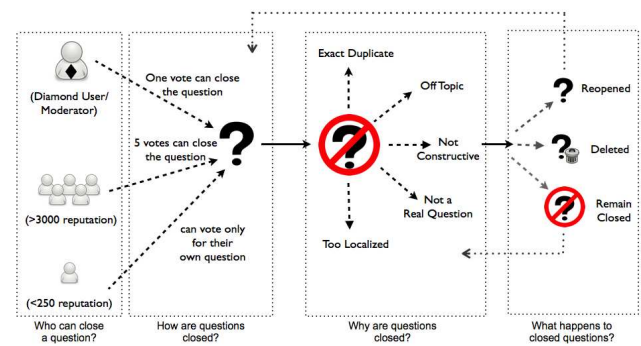}
\caption{depicts  \emph{who}, \emph{how} and \emph{why} questions are marked `closed' on Stack Overflow.}
\label{fig:so-diag}
\end{figure}

In this section, we discuss details on \emph{who}, \emph{how} and \emph{why} questions are closed on Stack Overflow. We also briefly outline \emph{what happens} once a question is `closed' and mention the community process rules to mark a question as `closed'. Figure~\ref{fig:so-diag} summarizes the details of important aspects of `closed' questions on Stack Overflow. 

\paragraph{What is a `closed' a question?}

A question can be `closed' on Stack Overflow if it is deemed unfit for its Q\&A format~\cite{:2012zr}. A `closed' question can not be answered but edits on previously posted question-answers and comments are permitted (subject to appropriate edit privileges). Question-answers can also be voted upon and are counted towards reputation points of users as well as badges. 

\paragraph{Who can `close' a question?}

Experienced users and community moderators can cast a vote to `close' a question. Stack Overflow users with 3,000+ reputation points and community moderators (also called $\Diamondblack$ moderators) can vote for the same. In addition, users with at least 250 reputation points can vote to `close' their own question. The \emph{Who} block of Figure~\ref{fig:so-diag} corresponds to the aforementioned details.

\paragraph{How are questions `closed' ?}

A question is automatically marked `closed' if it receives 5 `close' votes. However, $\Diamondblack$ moderator `close' votes are final and binding i.e. if a $\Diamondblack$ moderator decides to cast a `close' vote the question is `closed' immediately~\cite{:2012ly}. One can only vote once to `close' a question. The \emph{How} block of Figure~\ref{fig:so-diag} corresponds to this process.

\paragraph{Why are questions `closed'?}
According to Stack Overflow guidelines, a question is `closed' on Stack Overflow if it falls into one of the following five categories~\cite{:2012zr}: 

\begin{enumerate}
	\item \textbf{Exact Duplicate} -- contains similar content to previously posted questions
	\item \textbf{Off Topic} -- unrelated to programming scope as defined by Stack Overflow
	\item \textbf{Subjective (Not Constructive)} -- more likely to generate debates, discussions instead of answers
	\item \textbf{Not a Real Question} -- ambiguous, vague questions which do not have answers 
	\item \textbf{Too Localized} -- relevant to a very small geographic location, software or community
\end{enumerate} 

The \emph{Why} block of Figure~\ref{fig:so-diag} corresponds to this section.

\paragraph{What happens to a `closed' question?}

A `closed' question can be `reopened' if the question is improved from its current form. The `reopen' voting procedure is similar to the `close' procedure. However, if the questions are very poor in quality and beyond improvement, then they are \emph{deleted} from Stack Overflow~\cite{:2012zr}. The \emph{What} block of Figure~\ref{fig:so-diag} corresponds to this section.

\section{Characterization Study of `Closed' Questions}
\label{sec:clo-q-analysis}

In the first part of our work, we perform a characterization study of `closed' questions on Stack Overflow.

\subsection{Dataset Description}

Stack Overflow provides all user-generated content on its website for download under the \emph{Creative Commons Attribute-ShareAlike} license \cite{Atwood:2009uq}. We download Stack Overflow website data from the Stack Exchange August 2012 data dump provided by Stack Overflow which contains all data between July 31st, 2008 (the genesis of Stack Overflow) to August 31st, 2012 \cite{Overflow:2012yq}. Table~\ref{tab:data-stats} outlines basic statistics for Stack Overflow August 2012 dataset used in our characterization study. The statistics show that Stack Overflow is a very popular programming CQA with 1.29M registered users, 3.4M questions and 6.8M answers. 

\begin{table}[ht!] \scriptsize
\captionsetup{font=footnotesize, labelfont=bf, textfont=bf}
\centering
\caption{Stack Overflow August 2012 dataset statistics}
\begin{tabular}{l|l} \hline
Users & 1.29M (625k askers, 443k answerers) \\ 
Questions & 3.4M (62.21\% with accepted answers) \\ 
Answers & 6.8M  (31.33\% marked as accepted) \\ 
Votes & 27.5M  (72.35\% positive, 6.81\% favorites)\\ 
Ratio of Answers   & \multirow{2}{*}{2.16} \\
to Questions & ~  \\
\hline
\end{tabular}
\label{tab:data-stats}
\end{table}

In this work, we concentrate on `closed' questions on Stack Overflow i.e. questions which are deemed unfit and therefore low quality given the context. We extract all questions from the dataset which have been marked `closed' at least once. We find that approx. \textbf{3\% (0.1 Million)} questions are marked `closed' on Stack Overflow between August 2008 to August 2012. We use this data of \textbf{102,993} `closed' questions to conduct our characterization study and report our findings. Table~\ref{tab:cloq-data-stats} contains details on `closed' questions in Stack Overflow.~\footnote{\scriptsize{Prior to June 2011, `Close Votes' expired 4 days after their cast and are deleted from the dataset published by Stack Overflow. This information is available only if a question is closed successfully.}}

\begin{table}[ht!] \scriptsize
\captionsetup{font=footnotesize, labelfont=bf, textfont=bf}
\centering
\caption{Statistics of `Closed Questions' in Stack Overflow from August 2008 to August 2012.}
\begin{tabular}{l|l|l|l|l|l|l} \hline
~ & 2008 & 2009 & 2010 & 2011 & 2012 & Total \\ \hline
Closed & 3.8\%  & 1.52\%  & 1.77\% & 3.33\% & 3.82\% & 102, 993 \\ 
Questions & ~ & ~ & ~ & ~ & ~ & (2.98\%) \\  \hline
Closed& 0.03\%$^3$ & 0.25\%$^3$ & 0.75\%$^3$ & 2.21\% & 3.9\% & 570,418$^3$ \\
Votes &  &  &  & ~ & ~ & (0.2\%) \\  \hline
Ratio of  & ~ & ~ & ~ & ~ & ~ &  ~ \\
Answers to & 8.0 & 5.93 & 3.11 & 1.92 & 1.55 & 1.92  \\  
Questions & ~ & ~ & ~ & ~ & ~ &  ~ \\
\hline
\end{tabular}
\label{tab:cloq-data-stats}
\end{table}

\begin{figure}[ht!]
\captionsetup{font=footnotesize, labelfont=bf, textfont=bf}
\centering
\includegraphics[width=\linewidth]{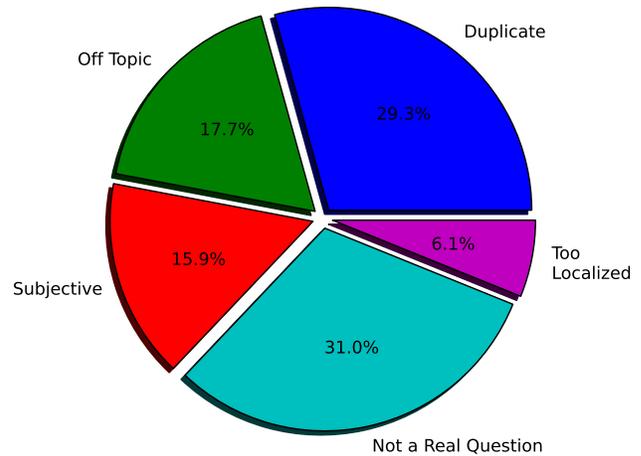}
\caption{shows the distribution of all five sub-categories of closed questions in our dataset.}
\label{fig:close-reason-distr}
\end{figure}

Based on the data, we can make two observations -- (1) Stack Overflow maintains a very good signal-to-noise ratio as reported in previous work~\cite{mamykina2011design} and (2) Despite the presence of vibrant community and structured guidelines, users do post questions which are unfit for the website.  A question can be closed on Stack Overflow for five reasons -  \emph{duplicate, off-topic, subjective, not a real question} and \emph{too localized}. Figure~\ref{fig:close-reason-distr} shows a pie-chart which depicts the distribution of `closed' questions on different sub-categories or reasons.~\emph{Not a Real Question} and \emph{Duplicate} categories are the most common reasons to close a question while \emph{Too Localized} is the least common reason.

\subsection{Temporal Distribution Analysis}

\begin{figure}[ht!]
\captionsetup{font=footnotesize, labelfont=bf, textfont=bf}
\centering
\includegraphics[width=\linewidth]{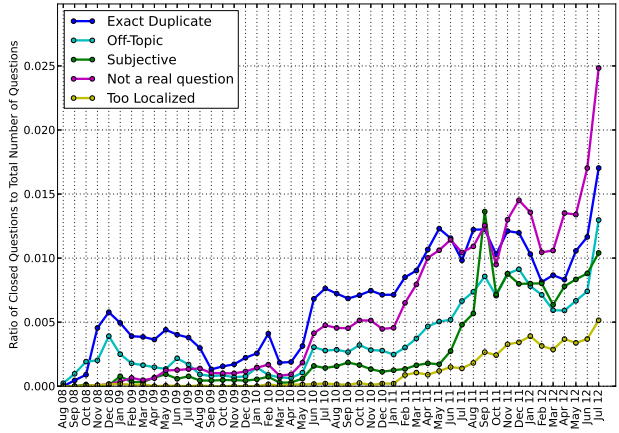}
\caption{shows the temporal distribution plot of the ratio of `closed questions' to total questions over a 48-month period between August 2008 to August 2012 for each sub-category. }
\label{fig:close-temporal-distr}
\end{figure}

We analyze the presence of `closed' questions on Stack Overflow over a 48-month time window between August 2008 to August 2012. Figure~\ref{fig:close-temporal-distr} depicts the ratio of `closed' questions to total questions over this time period. Overall, we find an increasing trend of the percentage of `closed' questions in each category i.e. we find that the number of questions `closed' over time has an upward curve. We also see that the most common categories of `closed' questions over 48-months are \emph{Exact Duplicate} and \emph{Not a Real Question}. Both these categories dominate in presence over the others across time. We perform qualitative analysis of some sample questions in our dataset to understand this pattern. The high ratio of the \emph{Exact Duplicate} category may be due to the problem of question retrieval on Stack Overflow i.e. users are unable to efficiently locate questions which are similar to the actual problem they are faced with. The presence of such a high ratio may also be due to lethargic users who do not perform adequate searches before posting a question. Similarly, the high percentage of \emph{Not a Real Question} category may be due to newly registered users who are yet to understand the scope, structure and guidelines of Stack Overflow. Overall, we see a sharp increase in the ratio of `closed' questions after January 2011.  

\subsection{Effect of New Registered Users}

\begin{figure}[ht!]
\captionsetup{font=footnotesize, labelfont=bf, textfont=bf}
\centering
\includegraphics[width=\linewidth]{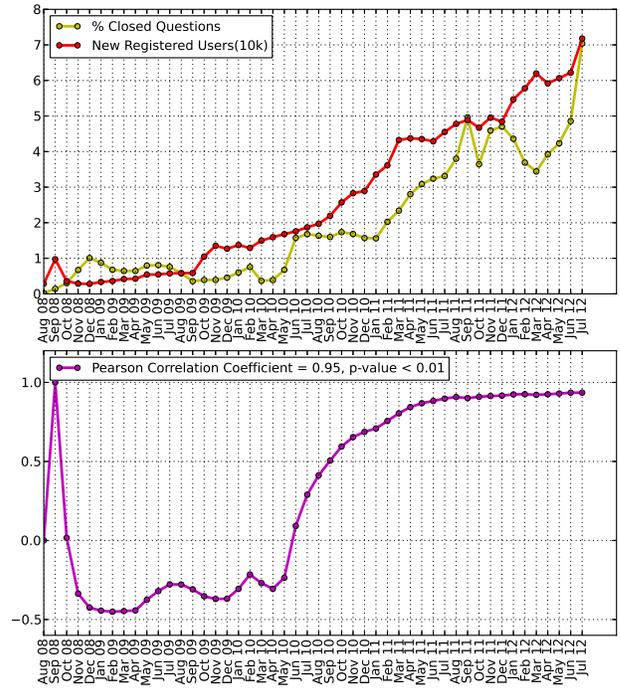}
\caption{shows the temporal distribution plot of the percentage of `closed questions' and newly registered users over a 48-month period from August 2008 to August 2012. In addition, the figure also shows correlation between both distributions.}
\label{fig:close-user-corr}
\end{figure}

Questions are marked `closed' on Stack Overflow if they are considered unfit for its Q\&A format. Intuitively, newly registered users on the website may be indolent to existing guidelines and may ignore them in their anxiety to get a solution to a problem. Therefore, we try to understand the impact of newly registered users on the presence of `closed' questions on Stack Overflow over time. Figure~\ref{fig:close-user-corr} shows the distribution of -- (1) number of newly registered users and (2) percentage of `closed' questions on Stack Overflow -- over a 48-month period between August 2008 to August 2012. In addition, it also depicts the corresponding \emph{Pearson Correlation Coefficient} (PCC) between the two distributions (cumulative) at each time interval. PCC calculates the linear dependence between two distributions and outputs a value between +1 (positive correlation) to -1 (negative correlation). Figure~\ref{fig:close-user-corr} shows a high correlation between the number of newly registered users and percentage of closed questions. We stress that the calculated correlation coefficient is between new registered users and the \textbf{percentage of closed questions} (and not the total number of closed questions) over time. The PCC value is +0.95 which indicates a very high correlation between the distributions with an extremely high confidence interval (p-value $<$ 0.01). The PCC shows that newly registered users may have an immediate impact on low quality content. Here too, we find a sharp rise in PCC after January 2011.

\subsection{Community Participation}

\begin{figure}[ht!]
\captionsetup{font=footnotesize, labelfont=bf, textfont=bf}
\centering
\includegraphics[width=\linewidth]{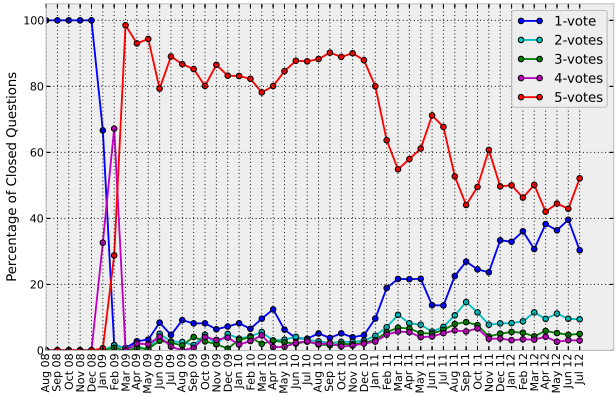}
\caption{shows the temporal distribution of `close votes' in closed questions over a 48-month period from August 2008 to August 2012. We observe that a high percentage of questions are closed due to a single $\Diamondblack$ moderator vote.}
\label{fig:close-vote-temporal}
\end{figure}

Stack Overflow follows a well defined community based voting procedure to evaluate a question before closure. We analyze these voting patterns to understand \emph{community participation} of experienced users and community moderators to weed out low quality content on the website. We recall that users with 3,000+ reputation points and $\Diamondblack$ moderators can cast a vote to close a question. A question is automatically `closed' if it reaches 5 votes but a vote from a $\Diamondblack$ moderator is binding and hence, immediately closes a question. Therefore, a question can be closed with any number of `close' votes between 1 to 5. Figure~\ref{fig:close-vote-temporal} shows the temporal distribution of `close' votes on Stack Overflow between August 2008 to August 2012. Table~\ref{tab:close-fav-vote-distr} (\emph{Close Votes} column) shows the distribution of number of `close votes' on closed questions. A significant percentage($\approx$40\%) of questions are closed due to a single $\Diamondblack$ moderator vote. More than 50\% of questions require $\Diamondblack$ moderator intervention to close a question. We also observe a rise in the percentage of questions being closed only by $\Diamondblack$ moderators over time. Simultaneously, we see a decrease in percentage of questions being closed by experienced users viz. users with 3,000+ reputation points. This shows that community participation to close questions is on a decline which has led to an increase in work load for $\Diamondblack$ moderators on this front. A $\Diamondblack$ moderator on Stack Overflow has also confirmed an increase in moderation work load over the years~\cite{:2013kx}. Stack Overflow has only 16 $\Diamondblack$ moderators for their website out of which 13 have been elected and 3 have been appointed~\cite{:2013ys}.

\begin{figure}[ht!]
\captionsetup{font=footnotesize, labelfont=bf, textfont=bf}
\centering
\includegraphics[width=\linewidth]{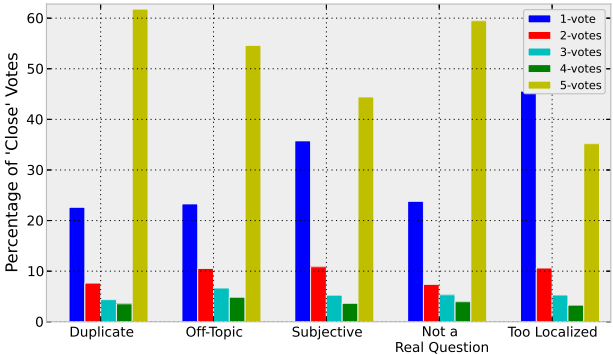}
\caption{shows the `close vote' distribution for each sub-category for all closed questions between August 2008 to August 2012. At least 1 out of 5 questions in each category are closed by a single $\Diamondblack$ moderator vote.}
\label{fig:close-vote-distr}
\end{figure}

We now analyze the  `close vote' patterns across each category of closed questions. Figure~\ref{fig:close-vote-distr} shows the `close vote' distribution for each sub-category of closed questions on Stack Overflow between August 2008 to August 2012. We see a strong community participation on \emph{Duplicate}, \emph{Off Topic} and \emph{Not a Real Question} categories. On the other hand, \emph{Subjective} and \emph{Too Localized} categories require a high amount of $\Diamondblack$ moderator intervention. We argue that the community participation behavior may be so because \emph{Duplicate}, \emph{Off Topic} and \emph{Not a Real Question} questions are low hanging fruits and easy to detect. The \emph{Subjective} category sees an equal community and $\Diamondblack$ moderator participation. The \emph{Too Localized} category sees a higher $\Diamondblack$ moderator intervention. Since, the presence of this category is very low in our dataset, such behavior may be primarily due to low traction owing to the difficulty of identification of such questions during normal daily usage of the website.

\subsection{Content Analysis}
\label{sec:content-analysis}

We now characterize the content of `closed' questions on Stack Overflow based on question tittle, question content, code snippets and topics. 

\paragraph{\textbf{Question Title, Body and Code Snippet}}

\begin{figure}[ht!]
\captionsetup{font=footnotesize, labelfont=bf, textfont=bf}
\centering
\includegraphics[width=\linewidth]{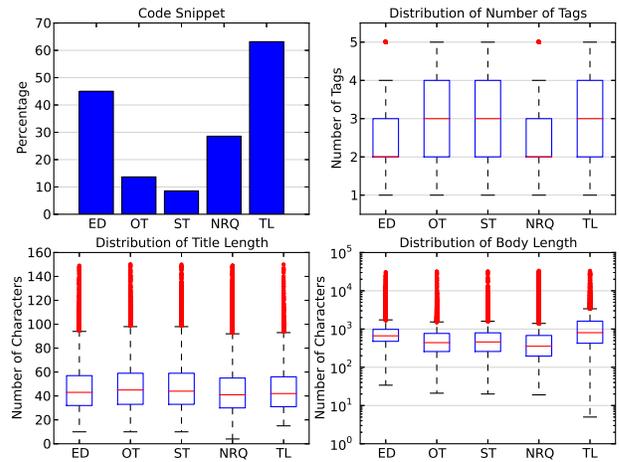}
\caption{shows the percentage of code snippets in each sub-category and character length distributions of question title, body as well as the distribution of number of tags in form of a box-and-whisker plot.(ED = Exact Duplicate, OT = Off-Topic, ST = Subjective, NRQ = Not a Real Question, TL = Too Localized)}
\label{fig:multi-box}
\end{figure}

Since, `closed' questions are unfit for Stack Overflow -- the presence or absence of code snippets may reveal insights about `closed' questions. Overall, $\approx$ 31\% of `closed' questions contain code snippets and hence, questions are `closed' even if they contain source code.  We analyze the presence of code snippets across each category to check if there are relative differences across categories. Figure~\ref{fig:multi-box} (left-top) shows the percentage of questions which contain code snippets for each category. We find that \emph{Too Localized} and \emph{Exact Duplicate} category contains a large number of questions which have code snippets in them. The \emph{Exact Duplicate} category by definition contains duplicate information to an existing question which may explain the high number. On the other hand, \emph{Too Localized} category by definition contains questions which are programming-related but are confined to a small community and hence, the higher percentage of questions which contain source code. We see that the \emph{Subjective} category contains the lowest percentage of questions containing source code. This could be probably because questions in this category are open-ended and invite discussions rather than an answer to a specific problem.

Figure~\ref{fig:multi-box} also shows the character length distribution of question title, body as well as the distribution of number of tags in form of a box-and-whisker plot. The top-right box plot shows that questions in the \emph{Exact Duplicate} and \emph{Not a Real Question} categories have lesser number of tags associated with it. The \emph{Exact Duplicate} category may exhibit such a behavior due to user lethargy while questions belonging to the \emph{Not a Real Question} category may be so as by definition the question marked with this label are non-programming related. A minimum of 1500 reputation points are required to create new tags on Stack Overflow~\cite{:vn}. The bottom left and bottom right box plots show the distribution of question title and question body lengths respectively. We do not observe a major difference in the length distributions either in title or body between categories. Both distributions are skewed i.e. there are may outliers (red points on the box plot) and the medians are approximately similar. However, in both of these distributions we once again see that the \emph{Not a Real Question} has the lowest median value which indicates that questions belonging to this category are a clear misfit to the Stack Overflow Q\&A format even in terms of content. 

\paragraph{\textbf{Question Topics}}
Each Stack Overflow question has some \emph{tags} associated with it which is an identification of the topic of the question content. We analyze frequently occurring tags in `closed' questions and bucket them into categories. Table~\ref{tab:cloq-tags-popular} shows popular tags in `closed' questions according to different categories. We see that popular tags on `closed' questions are similar to those found overall on Stack Overflow.

\begin{table}[ht!] \scriptsize
\captionsetup{font=footnotesize, labelfont=bf, textfont=bf}
\centering
\caption{Popular Tags in Closed Questions}
\begin{tabular}{l|l} \hline  
\textbf{Type} & \textbf{Tags} \\ \hline
Languages & java, c++, python, c, perl, r, $\hdots$ \\
Web2.0 & php, html5, html, css, apache, javascript, $\hdots$ \\
Operating Systems & iOS, unix, linux, android, ubuntu, osx, windows, $\hdots$ \\
Social & Facebook, wordpress, google, $\hdots$ \\
Miscellaneous & books, interview-questions, fun, homework, $\hdots$ \\
\hline
\end{tabular}
\label{tab:cloq-tags-popular}
\end{table}

We now analyze if `closed' questions contain certain topics which are unique to their category viz. tags which relatively occur more frequently in `closed' questions than otherwise. In order to do so, we normalize the occurrence of tags in `closed' questions by calculating the \textbf{Normalized Tag Ratio (NTR)} for each tag. Let $CQ$ be the set of all `closed' questions in Stack Overflow data, $NCQ$ be the set of all non-`closed' questions, $t_i$ be the set of tags in $CQ$ and $t_j$ be the set of tags in $NCQ$. We add the $\epsilon$ factor for smoothing purposes. Then,

\begin{math}
\forall t_i \quad in \quad CQ \quad where \quad t_i \in \{t_1 \dots t_n\}, \quad R_{CQ}^{i} = \frac{count(t_i)}{\sum_{i=1}^{n} count(t_i)} \\
\forall t_j \quad in \quad NCQ \quad where \quad t_j \in \{t_1 \dots t_m\}, \quad R_{NCQ}^{j} = \frac{count(t_j)}{\sum_{j=1}^{m} count(t_j)} \\
\end{math}
$$\therefore \forall t_i \quad in \quad CQ \quad NTR_{t_i} = \frac{R_{CQ}^{i}}{R_{NCQ}^{i} + \epsilon}$$

\begin{figure}[ht!]
\captionsetup{font=footnotesize, labelfont=bf, textfont=bf}
\includegraphics[width=\linewidth]{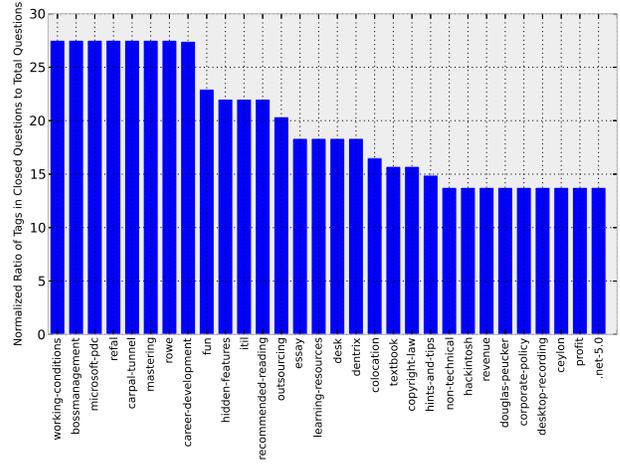}
\caption{shows the tags of closed questions on Stack Overflow with top 30 Normalized Tag Ratios (NTR).}
\label{fig:ratio-close-norm-tags}
\end{figure}

Figure~\ref{fig:ratio-close-norm-tags} shows the tags with top 30 $NTR$ in closed questions on Stack Overflow. We can now see tags which are unique to `closed' questions and find that these are quite different to the most popular tags. We notice that most tags are non-programming related; for example \emph{working-conditions, career-development, fun} etc. We also notice that some of these tags are programming related but are on broad topics like \emph{hidden-features, hints-and-tips} and \emph{textbook}. These tags are usually attached to questions which require a discussion and may not focus on problem specific solutions.

\subsection{Community Value and Information Quality}

A `closed' question is irrelevant to the Q\&A format and hence, implicitly suggests that the question may be low quality in context of Stack Overflow. Here, we analyze different indicators of content quality like \emph{Favorite Votes}, \emph{Closure Time}, \emph{Question Scores and Answering Patterns} and \emph{Question Status} with respect to `closed' questions. \\

\paragraph{\textbf{Favorite Votes}}

Stack Overflow provides its users a feature to \emph{favorite} a question. A \emph{favorite vote} is an explicit statement of approval by the user that she finds the question useful and appropriate. Table~\ref{tab:close-fav-vote-distr} (\emph{Favorite Votes} column) shows the cumulative distribution of `favorite votes' on overall closed questions. The data shows that $\approx$ 19\% of the overall `closed' questions receive at least one \emph{favorite vote} while $\approx$ 3\% of those receive $\geq$ 5 \emph{favorite votes}. 

\begin{table}[ht!] \scriptsize
\captionsetup{font=footnotesize, labelfont=bf, textfont=bf}
\centering
\caption{`Close Vote' Distribution and `Favorite Vote' Cumulative Distribution in Stack Overflow data. 45\% questions have at least one $\Diamondblack$ moderator vote and 26.5\% of questions are closed by a single $\Diamondblack$ moderator vote.  Approximately 1 out of 5 `closed' questions have at least 1 `favorite' vote and 3\% have at least 5 `favorite' votes.}
\begin{tabular}{l|l|l|l} \hline
\multicolumn{2}{c}{\textbf{Close Votes}} & \multicolumn{2}{c}{\textbf{Favorite Votes}} \\  \hline
\textbf{Votes} & \textbf{Closed Questions} & \textbf{Votes} & \textbf{Closed Questions} \\ \hline
1-vote & 27,390 (26.59\%) & $\geq$ 1 & 19,156(18.6\%) \\ 
2-votes & 9,037 (8.77 \%)  & $\geq$ 5 & 3,374(3.28\%) \\ 
3-votes & 5,436 (5.28\%)  & $\geq$ 10 & 1,872(1.82\%)   \\
4-votes & 4,030 (3.91\%)  & $\geq$ 100 & 206(0.2\%)  \\
5-votes & 57,117 (55.44 \%) & $\geq$ 500 & 29(0.03\%) \\ \hline
Total & 102,993 & ~ & ~ \\
\hline
\end{tabular}
\label{tab:close-fav-vote-distr}
\end{table}

\begin{figure}[ht!]
\captionsetup{font=footnotesize, labelfont=bf, textfont=bf}
\includegraphics[width=\linewidth]{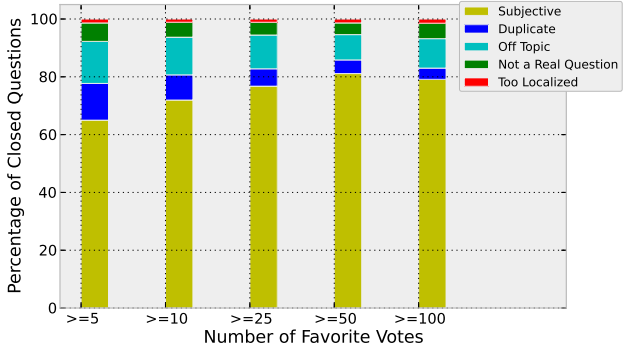}
\caption{shows the distribution of  `favorite votes' on closed questions for each sub-category on various thresholds. \emph{Subjective} category attracts very high number of \emph{favorite votes} from users.}
\label{fig:cloq-fav-reason}
\end{figure}

However, features such as \emph{likes} and \emph{favorite votes} are known to be abused by users for purposes other than their intended use. Therefore, we analyze \emph{favorite vote} distributions on different thresholds for all sub-categories of closed questions. Figure~\ref{fig:cloq-fav-reason} shows the distribution of \emph{favorite votes} at different thresholds for each category of `closed' questions. We see that the \emph{Subjective} category attracts a very high number of \emph{favorite votes} from users. We perform a manual qualitative analysis on these questions and notice that the \emph{Subjective} category contains questions like Polls, Hidden Features, Books, Tricks, Interview Questions and Open ended questions. Table~\ref{tab:sample-fav-vote-qns} shows examples of questions in the \emph{Subjective} category which have $\geq$ 100 \emph{favorite votes}. Note that our analysis in Section~\ref{sec:content-analysis} showed that \emph{Subjective} category had the lowest percentage of questions containing code snippets. Therefore, despite the emphasis on objectivity and source code related questions by Stack Overflow guidelines we see that some amount of programming related \emph{Subjective} questions are encouraged and appreciated by the community members. 

\begin{table}[ht!] \scriptsize
\captionsetup{font=footnotesize, labelfont=bf, textfont=bf}
\centering
\caption{Example questions with $\geq$ 100 `favorite votes' on closed questions in \emph{subjective} category.}
\begin{tabular}{l|l|l|l} \hline
\textbf{Favorites} & \textbf{Title} & \textbf{Answers} &\textbf{Views}\\ \hline
5894 & List of freely available programming books & 112 & 569,199\\
2228 & Hidden features of Python & 100 & 212,589 \\
\multirow{2}{*}{1685} & What is the best comment in source & \multirow{2}{*}{519}  & \multirow{2}{*}{1,051,784} \\
 & code you have ever encountered? & & \\
421 & Worst security hole you've seen? & 163  & 32,840 \\
140 & What is the most useful R trick? & 34 & 13,197\\
\hline
\end{tabular}
\label{tab:sample-fav-vote-qns}
\end{table}

\paragraph{\textbf{Closure Time}}

\begin{figure}[ht!]
\captionsetup{font=footnotesize, labelfont=bf, textfont=bf}
\includegraphics[width=\linewidth]{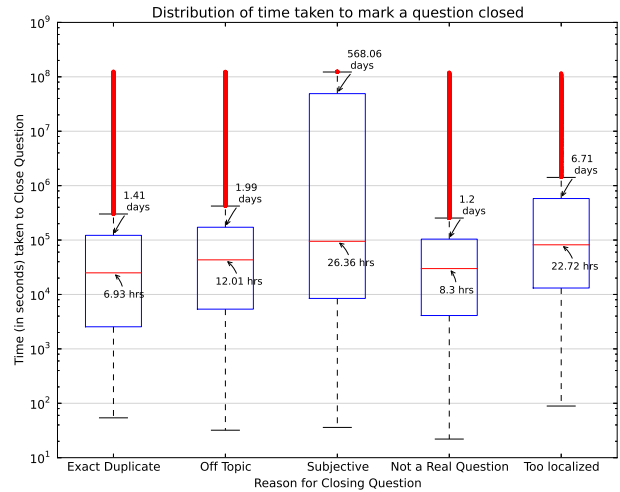}
\caption{shows the distribution of time taken to close questions for each category in the form of a box-and-whisker plot.}
\label{fig:timediff-close-ques}
\end{figure}

We now analyze the time taken to `close' questions on Stack Overflow. Figure~\ref{fig:timediff-close-ques} shows the closure time distribution of  `closed' question for every sub-category. The median closure times for \emph{Exact  Duplicate}, \emph{Off Topic} and \emph{Not a Real Question} is 6.93, 12.01 and 8.3 hours respectively. Most questions in these categories are quickly turned towards closure which may signify that their community value is relatively low than other categories. The \emph{Subjective} and \emph{Too Localized} categories have the highest median closure time $\approx$ 26 and 22 hours respectively. The reason for high closure time for the \emph{Subjective} category could be because most questions (despite not being a good fit) invite discussion and opinions on broad programming related principles, guidelines, polls etc. Therefore, it takes time before these questions are answered in entirety and hence are left open for a longer time. We also notice a higher \emph{spread} of closure times (upper quartile=586.06 days) in this category demonstrating that if a \emph{Subjective} question is not closed within 1 day it takes a long time to close the question. The community actions indicate that these questions have not reached their maximum community value potential and hence remain open. Our prior analysis for `close vote' distribution for \emph{Too Localized} category in Figure~\ref{fig:close-vote-distr} shows that the 66\% of the questions in this category requires moderator intervention which may be one of the reasons for higher closure time. Even though \emph{Too Localized} category has a similar median closure time (22.72 hours) to the \emph{Subjective} category very few questions require more than 6.71 days to close. This indicates that questions in the \emph{Too Localized} category, despite similar median closure time, reach their maximum community value potential relatively earlier than those in \emph{Subjective} category. We also find that each category contains some outliers i.e. each category contains some questions which take a long time to be marked as `closed'. Table~\ref{tab:close-vote-outliers} shows the close vote distribution pattern on questions with outlier closure times in each category.

We see that all the outlier questions have a very high percentage of $\Diamondblack$ moderator intervention on question closure time. This indicates that these questions are indeed outliers in terms of content too as the community prefers to keep these questions open to reach their maximum community value potential.

\begin{table}[h] \scriptsize
\captionsetup{font=footnotesize, labelfont=bf, textfont=bf}
\centering
\caption{Number of Close Votes on outliers from each category}
\begin{tabular}{llllll} \hline
\textbf{Category} & \textbf{1-vote} & \textbf{2-vote}  & \textbf{3-vote} & \textbf{4-vote} & \textbf{5-vote} \\ \hline
Duplicate & \textbf{55.44\%} & 11.68\% & 4.25\% & 2.18\% & 26.45\% \\
Off-Topic & \textbf{42.06\%} & 16.21\% & 6.31\% & 3.47\% & 31.96\% \\
Subjective & \textbf{64.64\%} & 16.66\% & 4.9\% & 2.26\% & 11.54\% \\
Not a Real Question & \textbf{46.97\%} & 9.52\% & 6.28\% & 3.5\% & 33.74\% \\
Too Localized & \textbf{68.22\%} & 11.85\% & 3.62\% & 1.86\% & 14.45\% \\
\hline
\end{tabular}
\label{tab:close-vote-outliers}
\end{table}

\paragraph{\textbf{Question Scores and Answer Patterns}}

\begin{figure}[ht!]
\captionsetup{font=footnotesize, labelfont=bf, textfont=bf}
\centering
\includegraphics[width=\linewidth]{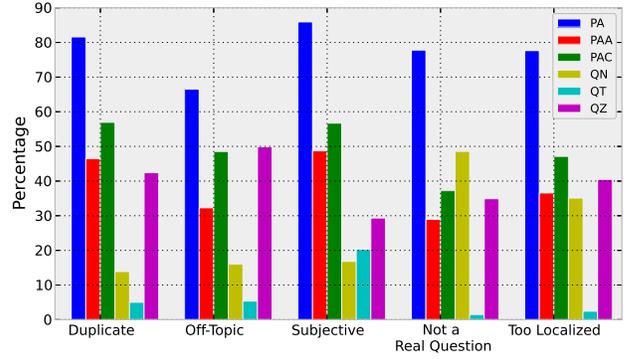}
\caption{shows the question scores and answering patterns of users on closed questions in each category. PA = Percentage of Answers, PAA = Percentage of Accepted Answers, PAC = Percentage of Accepted Answers given that a `closed' question has an answer, QN = Percentage of Questions with Negative Score, QT = Percentage of Questions with $\geq$ `t' Score (t=5), QZ = Percentage of Questions with Zero Score.}
\label{fig:answer-patterns}
\end{figure}

Figure~\ref{fig:answer-patterns} shows various question scores and answer patterns on `closed' questions in Stack Overflow. We first look into the percentage of answers (PA), percentage of accepted answers (PAA)\footnote{\scriptsize{A question asker can mark an answer \emph{accepted} if the answer solves the problem faced by the question asker. Accepted answers lead to gain in reputation points to answerers.}} and percentage of accepted answers given an answer (PAC) on each category of `closed' questions on Stack Overflow.  We see that a large percentage of `closed' questions receive answers from users. The \emph{Duplicate} and \emph{Subjective} categories also have a relatively higher PA and PAA than other categories. The higher PA and PAA on \emph{Duplicate} questions suggest that despite the fact that the question content is an exact duplicate of others the community is eager to answer the question. Such behavior may also be exhibited by answerers to garner more reputation points in the form of answer votes and \emph{accepted} answers. This may also explain why we see a very low PAA in the \emph{Not a Real Question} category as users are smart enough to pick questions which have a higher probability of receiving up votes. Recall that our earlier analysis reveals that  questions belonging to the \emph{Not a Real Question} category are low in information content quality. We also analyze question score patterns on each category of `closed' questions on Stack Overflow. We calculate percentage of questions with negative score (QN), percentage of questions with $\geq 5$ score (QT) and percentage of questions with zero score (QZ). We find that \emph{Not a Real Question} has the highest QN and once again indicates that questions in this category are very low in quality. We observe a similar pattern for \emph{Too Localized} category and may indicate that the community in general frowns upon questions which are too confined to certain sections of the programming fraternity. We see that \emph{Subjective} category has a very high QT and this falls in line with our earlier hypothesis that questions in this category despite being not fit to the website are immensely popular and therefore, draws large number of votes. We see similar QZ values (between 30--50\%) on all categories of `closed' questions which demonstrates that some questions do not get any approval from the community. We would like to comment that we do not observe any familiar statistical distributions like power-law on any of these question scores and answer patterns. 

\paragraph{\textbf{Question Status}}

Apart from being marked as `closed', a Stack Overflow question can also be given a \emph{locked}, \emph{community wiki} and \emph{protected} label. Table~\ref{tab:close-question-status} shows the distribution of `closed' questions with a \emph{locked}, \emph{community wiki} and \emph{protected} label. A \emph{locked} question can not receive any new answers or any form of votes on question-answers. A question is primarily \emph{locked} by $\Diamondblack$ moderator to prevent gaming or abuse of the system by users to garner reputation points~\cite{:2008kx}. We observe that \emph{Exact Duplicate} and \emph{Off Topic} categories are most prone to reputation gaming and therefore, marked as \emph{locked}. A \emph{community wiki} label is an intent to `donate' and transfers ownership of the question from the asker to the community. The goal of Stack Overflow is to be a knowledge base of programming information and therefore \emph{community wiki} posts play a significant role in achieving that goal~\cite{:2013fk}. We see that questions from the \emph{Subjective} category contain a high number of \emph{community wiki} donations. We hypothesize that this would be due to the nature of \emph{subjective} questions as these contain discussions, opinions on programming topics which may be ``never ending" (philosophical rather than factual). A \emph{protected} label is an intent to prevent noisy answers like ``Thank You", ``+1" from new users who may not understand the guidelines of the forum. A \emph{protected} label prevents newly registered users from answering these question~\cite{:2010uq}. Once again we see that a high percentage of questions from \emph{Subjective} category are marked as \emph{protected}. This demonstrates that \emph{Subjective} questions are very attractive and ``fun" questions to users although they may not fit into the Stack Overflow guidelines.

\begin{table}[ht!] \scriptsize
\captionsetup{font=footnotesize, labelfont=bf, textfont=bf}
\centering
\caption{shows the distribution of `closed' Questions in the Stack Overflow with labels \emph{locked, community wiki} and \emph{protected}.}
\begin{tabular}{llll} \hline
~  & \multicolumn{3}{c}{\textbf{Number of `Closed' Questions}} \\ \hline 
\textbf{Category} &  \textbf{Locked} & \textbf{Community Wiki} & \textbf{Protected} \\ \hline
\textbf{Exact Duplicate} &  732(33.8\%) & 160(9.9\%) & 36(10.3\%) \\
\textbf{Off Topic} &  \textbf{1180(54.5\%)} & 273(16.8\%) & 70(20.1\%) \\
\textbf{Subjective} &  188(8.7\%) & \textbf{978(60.3\%)} & \textbf{202(58\%)} \\
\textbf{Not Real Question} & 50(2.3\%) & 192(11.8\%) & 28(8\%) \\
\textbf{Too Localized} & 114(0.6\%) & 10(0.6\%) & 12(3.4\%) \\ \hline
Total & 2,264 & 1,613 & 348 \\ 
\hline
\end{tabular}
\label{tab:close-question-status}
\end{table}

\subsection{Characterization Summary}
We now summarize key findings from our characterization study -- 
\begin{itemize}
	\item We see an increasing trend in the percentage of `closed' questions over time -- in particular \emph{Exact Duplicate} and \emph{Not a Real Question} -- with a steep rise after January 2011
	\item We find a positive correlation with a high confidence value between new registered users and the percentage of `closed' questions
	\item We observe a decrease in community participation to mark a question as `closed' over time which has probably led to increase in work load for $\Diamondblack$ moderators
	\item Popular tags on `closed' questions are very similar to overall questions but tags unique to `Closed' questions are vague and non-programming related
	\item Questions from the \emph{Subjective} category do not follow the Q\&A format but are very popular and have high community value. They also take relatively longer time to be marked as `closed'. Questions from the \emph{Not a Real Question} category take least amount of time to be closed and are low in community value.
	\item Despite a very high percentage of presence of source code, questions in \emph{Too Localized} are not very popular in the community
	\item \emph{Exact Duplicate} and \emph{Off Topic} questions are relatively more attractive to reputation gamers
	
\end{itemize}

\section{Prediction of `Closed' Questions}
\label{sec:clo-q-prediction}

In the second part of our study, we build a predictive model to automatically detect a `closed' question on Stack Overflow. We formulate the prediction of `closed' questions on Stack Overflow as a binary classification task. 

\subsection{Features for Classification}

We investigate \textbf{19} features based on \emph{User Profile}, \emph{Community Process}, \emph{Question Content} and \emph{Textual Style} for our prediction task. Table~\ref{tab:feat-set} shows different categories of feature sets used by our system for `closed' question prediction. \emph{User Profile} features are based on user's participation activity while \emph{Community Process} features are based on Stack Overflow community contributions in the form of votes, accepted answers etc.~\emph{Question Content} features are calculated by extracting the content from questions and \emph{Textual Style} features characterizes the writing and posting style of the question asker. It is important to note that there may be other distinguishing features for `closed' questions (for example - answering patterns) but the aim of the study is to predict a `closed' question at its \emph{creation time}. Hence, we can not make user of these features for our predictive model. The reputation of the user at question creation time is an excellent feature by intuition however, this data is not made available by Stack Overflow. Therefore, we use \emph{Community Process} features to offset for this missing data. In addition, questions are routinely edited (title, body and tags) by experienced community users. However, there is no mechanism to get the original text of the question. All these factors make prediction of a `closed' question difficult and challenging. 

\begin{table}[ht!] \scriptsize
\captionsetup{font=footnotesize, labelfont=bf, textfont=bf}
\centering
\caption{shows the different categories of feature sets used for `closed' question prediction}
\begin{tabular}{clcl} \hline 
\textbf{Set} & \textbf{Category} & \textbf{Number} & \textbf{Features} \\ \hline
\multirow{2}{*}{$A$} & \multirow{2}{*}{\textbf{User Profile}} & \multirow{2}{*}{3} & \emph{age of account, badge score}, \\
~ & ~ & ~ & \emph{previous negative posts} \\ \hline
\multirow{2}{*}{$B$} & \textbf{Community} & \multirow{2}{*}{4} & \emph{post score, accepted answer score,} \\
~ & \textbf{Process} & ~ & \emph{comment score, favorite score} \\ \hline
\multirow{2}{*}{$C$} & \multirow{2}{*}{\textbf{Question Content}} & \multirow{2}{*}{3} & \emph{number of urls, number of} \\
~ & ~ & ~ & \emph{stack overflow urls, number of popular tags} \\ \hline
\multirow{5}{*}{$D$} & \multirow{5}{*}{\textbf{Textual Style}} & \multirow{5}{*}{9} & \emph{title length, body length, number of tags} \\
~ & ~ & ~ & \emph{number of punctuation marks, number of } \\ 
~  & ~ & ~ & \emph{short words, code snippet length, number of } \\ 
~  & ~ & ~ & \emph{spl. characters, number of lower case} \\ 
~  & ~ & ~ & \emph{characters, number of upper case characters} \\ 
\hline
\end{tabular}
\label{tab:feat-set}
\end{table}

While most of the features are self-explanatory, below we explain some of the higher order features below (calculated at time of question creation) --

\textbf{Badge Score (BS)}:\\
Let $\{b_1 \dots b_n\}$ be the badges earned by the user, then
$$BS = \displaystyle\sum_{i=0}^{n}{\frac{1}{\textrm{\#users who have }b_i}}$$ \\
\textbf{Post Score (PS)}:\\
Let $\{q_1 \dots q_n\}$ be the set of previous questions asked by the user and $\{a_1 \dots a_m\}$ be the set of previous answers posted by the user, then \\
$$PS = \displaystyle\sum_{i=0}^{n}{score(q_i)} + \displaystyle\sum_{j=0}^{m}{score(a_i)}$$
\textbf{Comment Score (CS)}:\\
Let $\{c_1 \dots c_n\}$ be the set of previous comments posted by the user, then\\
$$CS = \displaystyle\sum_{i=0}^{n}{score(c_i)}$$ 
\textbf{Favorite Score (FS)}:\\
Let $\{fq_1 \dots fq_n\}$ be the set of questions asked by the user which have been favourited and $\{fa_1 \dots fa_m\}$ be the set of answers posted by the user which have been favourited, then \\
$$FS = \displaystyle\sum_{i=0}^{n}{score(fq_i)} + \displaystyle\sum_{j=0}^{m}{score(fa_i)}$$
\textbf{Accepted Answer Score (AAS)}:\\
Let $\{aa_1 \dots aa_n\}$ be the set of answers posted by the user which have been accepted. We give an individual score of 15 to each accepted answer, therefore\\
$$AAS = \displaystyle\sum_{i=0}^{n}{15}$$ 
\textbf{Number of Popular Tags (\#PT)}:\\
Let $T = \{t_1 \dots t_n\}$ be the tags present in the question, and $PT = \{pt_1 \dots pt_m\}$ our pre-derived set of popular tags on Stack Overflow\footnote{\scriptsize{We obtain popular tags by calculating tag distribution of all tags in our dataset.}}, then\\
$$\#PT = \|T \cap PT\|$$

\subsection{Experimental Testbed, Setup and Classifier}

Stack Overflow contains 102,993 `closed' questions between August 2008 to August 2012. Out of these questions, 1302 questions do not have any information about the question asker. We ignore these questions and consider the remaining \textbf{101,691} `closed' questions as our positive class. The percentage of non-`closed' questions (negative class) is very high  (97\%) than `closed' questions (3\%) and therefore, leads to the formation of an imbalanced dataset. Learning with imbalanced data is a research challenge and has attracted wide spread attention of researchers in the machine learning community. Various approaches have been proposed in literature to address the nature of imbalanced datasets. One such approach is to randomly under-sample the majority class data or over-sample the minority class data to make the dataset balanced~\cite{he2009learning}. In order, to make our dataset balanced we under-sample the majority class (non-`closed' questions or -ve class) and draw \textbf{101,691} random samples. However, random sampling may result in sample bias and lead to loss of information. In order to eliminate this sample bias, we perform under-sampling by drawing several random independent subsets from the majority class (-ve class) and training multiple classifiers based on each of these subsets along with the minority class (+ve class). We then evaluate our classifier across these multiple data instances and report our results. 

In our experiments, we draw 10 independent random subsets from \textbf{101,691} samples from the non-`closed' questions (negative majority class) and train 10 classifiers based on each of these 10 subsets along with \textbf{101,691} samples from `closed' questions (positive minority class). Therefore, in total we have \textbf{203,382} data samples across both classes for each classification run. We experiment with various classification algorithms including Support Vector Machines, Naive Bayes, Logistic Regression etc. and find that the \emph{Stochastic Gradient Boosted Trees} gives the best performance. \empty{Stochastic Gradient Boosted Trees} (SGBT) is an ensemble learning technique which combines information from \emph{weak} predictive models (primarily built on decision trees) to form a \emph{strong} classifier~\cite{friedman2002stochastic}. The stochastic approach randomly sub-samples the training data without replacement before the construction of each tree and hence, avoids over fitting on the data. Previous approaches in information and question quality prediction on CQA services have also observed good classification performance with SGBT~\cite{agichtein2008finding, Li:2012:APQ:2187980.2188200}. Table \ref{tab:exp-setup} provides a summary of our testbed and experimental setup. 
\begin{table}[ht!] \scriptsize
\captionsetup{font=footnotesize, labelfont=bf, textfont=bf}
\centering
\caption{Details of Experimental Setup}
\begin{tabular}{l|l} \hline 
\textbf{Dataset} & 203,382 questions\\ 
\textbf{`Closed' (+ve class)} & 101,691 \\
\textbf{Non-`Closed' (-ve class)} & 101,691 (drawn randomly from -ve class, 10 times)\\
\textbf{Classifier} &  Stochastic Gradient Boosted Trees (SGBT) \\ 
\textbf{Learning Rate} & 0.1 \\
\textbf{Sub-sample size} & 0.5 \\
\textbf{Classification Runs} & 10 (for each +ve/-ve random sample pair) \\
\textbf{Feature Sets} & \{A\}, \{A, B\}, \{A, B, C\}, \{A, B, C, D\} \\
\textbf{Train-Test Split} & 70\%-30\% \\
\textbf{Cross Validation} & 10-folds \\
\hline
\end{tabular}
\label{tab:exp-setup}
\end{table}

We choose standard parameters for learning rate (0.1) and sub-sample size (0.5) for SGBT. We use a 70-30\% training-testing split and perform 10-fold cross validation on each classification run of our positive class versus random sample of negative class. 

\subsection{Classification Results and Evaluation}

Table~\ref{tab:conf-mat} shows the confusion matrix for our classification experiments. We are able to accurately classify 76.5\% of `closed ' questions and 69.1\% of non-`closed' questions. 

\begin{table}[ht!] \scriptsize
\captionsetup{font=scriptsize, labelfont=bf, textfont=bf}
\centering
\caption{Confusion Matrix -- Classification Results}
\begin{tabular}{cccc} \hline
~ & ~ & \multicolumn{2}{c}{\textbf{Predicted}} \\ \hline
~ & ~ & \textbf{Closed} & \textbf{Non-Closed} \\
\multirow{2}{*}{\textbf{True}} & \textbf{Closed} & \textbf{76.5\%} & 23.5 \% \\
~ & \textbf{Non-Closed} & 30.9\% & \textbf{69.1\%} \\
\hline
\end{tabular}
\label{tab:conf-mat}
\end{table}

\begin{figure}[ht!]
\captionsetup{font=footnotesize, labelfont=bf, textfont=bf}
\centering
\includegraphics[width=\linewidth]{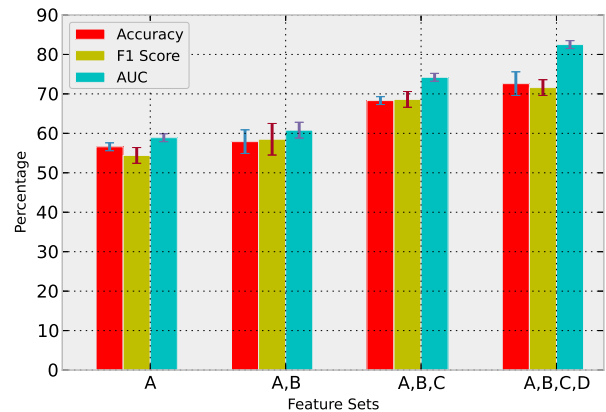}
\caption{shows classifier performance with Accuracy, F1 and Area Under the ROC curve (AUC) metrics when feature sets are incrementally added. Note the strong performance of our classifier on every feature set addition.} 
\label{fig:class-perf}
\end{figure}

Our characterization study indicates that there is no intuitive heuristic or metric to predict a `closed' question. Hence, in order to understand the effect of features to predict `closed' questions, we incrementally add feature sets to our classifier and record the performance. We use three standard information retrieval metrics -- F1 score, Accuracy and Area Under the ROC curve (AUC) to evaluate our classifier. Figure~\ref{fig:class-perf} shows the performance of our classifier on Accuracy, F1 score and AUC metrics when feature sets are incrementally added. We see that each feature set has a positive effect on the performance of the classifier across all metrics. This suggests that the all our feature sets are important for prediction.

\subsection{Feature Importance}

\begin{figure}[ht!]
\captionsetup{font=footnotesize, labelfont=bf, textfont=bf}
\centering
\includegraphics[width=\linewidth]{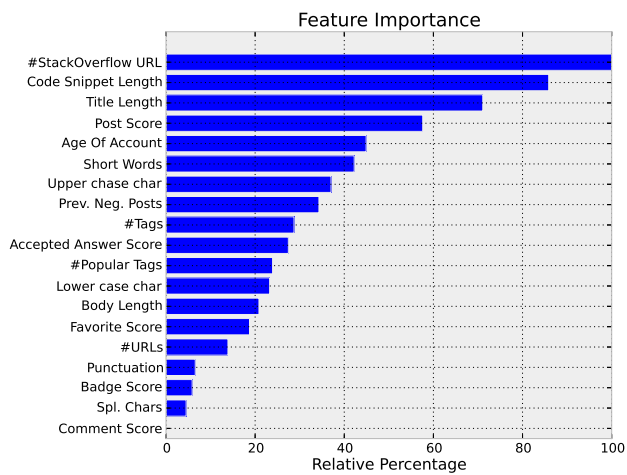}
\caption{shows the relative feature importance of all 19 features in our predictive model. }
\label{fig:imp-feat}
\end{figure}

One of the advantages of using SGBT is that it outputs a list of important features used for classification. Figure~\ref{fig:imp-feat} shows the most important features for classification. Overall, we see that almost all features contribute towards our prediction model. The \emph{Comment Score}, \emph{Special Characters}, \emph{Badge Score} and \emph{Punctuation Marks} features have little or no significance in our classifier. The top five features for classification are -- \emph{Stack Overflow URLs}, \emph{Code Snippet Length}, \emph{Title Length}, \emph{Post Score} and \emph{Age of Account}. The most important feature is the presence of \emph{Stack Overflow URLs} which suggests that good quality questions contain references to information within the website. \emph{Code Snippet Length} is also a differentiating factor which shows that despite presence of source code in `closed' questions the length of the code snippet is relatively low to that of non-`closed' questions. The presence of \emph{Title Length} as a top feature indicates that overall `closed' questions contain shorter titles than non-`closed' questions. \emph{Post Score} and \emph{Age of Account} points out that new users have a higher chance of question closure and probably they do not understand the Stack Overflow format.

\section{Conclusion}
\label{sec:clo-q-conclusion}

Stack Overflow is an extremely popular programming Community Question Answer (CQA) website for developers throughout the world. Stack Overflow uses a \emph{karma} based incentive system to maintain the quality of content on its website. However, despite these guidelines users post questions which do not fit Stack Overflow's Q\&A format. Questions which are deemed unfit for Stack Overflow are marked as `closed' by experienced users and community moderators. We present the first study of `closed' question on 4 years of publicly available data from Stack Overflow. We divide our study into two phases -- In the first phase, we conduct a characterization of `closed' questions posted between August 2008 to August 2012. Our characterization reveals that \emph{subjective} `closed' questions are popular and high quality while \emph{not a real question} are low in quality. We also notice decrease in community participation on question closure over time and find that \emph{Duplicate} and \emph{Off Topic} are more prone to reputation gaming. In the second phase, we construct a predictive model for identifying a `closed' question using an ensemble learning technique and report 73\% accurate predictions overall. Feature analysis reveals \emph{stack overflow urls} and \emph{code snippet length} as top differentiating features to predict a `closed' question.



%
\bibliographystyle{abbrv} \scriptsize
\bibliography{precloq}  
%
%
\end{document}